\documentclass[showpacs,twocolumn,superscriptaddress]{revtex4}
\usepackage{amsmath,amssymb,mathrsfs,graphicx}

\newcommand{\bs}[1]{\boldsymbol{#1}}
\newcommand{\comm}[2]{\left[#1,#2\right]}
\newcommand{\ket}[1]{\left|#1\right\rangle}
\newcommand{\bra}[1]{\left\langle#1\right|}
\newcommand{\ii}{\text{i}}

\allowdisplaybreaks[4]

\begin{document}
\title{Valence bond solid states with symplectic symmetry}  
\author{Dirk Schuricht}
\affiliation{The Rudolf Peierls Centre for Theoretical Physics, 
University of Oxford, 1 Keble Road, OX1 3NP, Oxford, United Kingdom}
\author{Stephan Rachel}
\affiliation{Institut f\"ur Theorie der Kondensierten Materie,
Universit\"at Karlsruhe, Postfach 6980, 76128 Karlsruhe, Germany}
\date{\today}

\pagestyle{plain}
\begin{abstract}
  We introduce a one-dimensional valence bond solid (VBS) state with
  symplectic symmetry SP($n$) and construct the corresponding parent
  Hamiltonian. We argue that there is a gap in the spectrum. We calculate
  exactly the static correlation functions, which fall off exponentially.
  Hence the model introduced here shares all properties of the Haldane
  scenario for integer-spin quantum antiferromagnets. We further show that the
  VBS state possesses string order and discuss its generalization to higher
  dimensions.
\end{abstract}
\maketitle

\section{Introduction}
In 1987 Affleck, Kennedy, Lieb, and Tasaki (AKLT)~\cite{Affleck-87}
introduced the SU(2) valence bond solid (VBS) state and showed that it
is the unique ground state of a special antiferromagnetic spin-1
chain.  This model possesses all properties of the Haldane scenario
for integer-spin quantum antiferromagnets~\cite{Haldane83pl1}, namely
a unique ground state, an energy gap between the ground state and the
excitations, and exponentially decaying correlations in the ground
state.  Soon after its discovery the VBS state was reformulated in
terms of Schwinger bosons~\cite{Arovas-88,Auerbach94}. This
formulation revealed a striking analogy between the VBS state and the
Laughlin state in the fractional quantum Hall effect~\cite{Laughlin83}
and enabled the analysis of the excitations above the AKLT state using
a single-mode approximation~\cite{Arovas-88}.  Following these
developments the AKLT model was widely used to study general
properties of spin-1 chains, for example the appearance of hidden
string order~\cite{denNijsRommelse89} and a
$\mathbb{Z}_2\times\mathbb{Z}_2$-symmetry breaking in the Haldane
phase~\cite{KennedyTasaki92prb}. This success has also motivated the
study of $q$-deformed AKLT chains~\cite{Kluemper-91} as well as
SU($n$) generalizations of the VBS
construction~\cite{Affleck-91,GRS07}.  Very recently, Tu {\it et al.}
extended the investigation of hidden string order to SO($n$) symmetric
Hamiltonians~\cite{Tu-08}.

Another important invention was the formulation of generalized VBS states in
terms of finitely correlated or matrix product
states~\cite{Fannes-91,Kluemper-93}, which in particular allows the relatively
easy calculation of correlation functions. \"Ostlund and
Rommer~\cite{OestlundRommer95} showed that the wave functions appearing in the
density matrix renormalization group (DMRG) method~\cite{White92} are
represented by matrix product states.  Since then DMRG algorithms, which make
direct use of the matrix product state formulation~\cite{Mcculloch07}, have
been developed.

On the other hand, large-$n$ techniques based on symplectic symmetry were
introduced by Read and Sachdev~\cite{ReadSachdev91} to study frustrated
antiferromagnets on a square lattice. If one places symplectic spins
transforming under a given representation of SP($n$) on an arbitrary lattice,
it is always possible to form singlet bonds between any two sites.  This is
not true for unitary spins transforming under SU($n$), where the formation of
singlets is in general only possible on bonds between a representation and
their complex conjugate representation. This restricts the applicability of
SU($n$) techniques to bipartite lattices.  The SP($n$) technique was
afterwards widely used to study frustrated antiferromagnets on various
lattices~\cite{Sachdev92}, doped antiferromagnets~\cite{Vojta-00}, paired
Fermi gases~\cite{SachdevWang91}, stripes in high-temperature
superconductors~\cite{VojtaRoesch08}, and heavy-fermion
systems~\cite{WeberVojta08}. Recently, Flint~\emph{et al.}~\cite{Flint-07}
introduced the ``symplectic-$n$'' approach which links time reversal and
symplectic symmetry of spins by eliminating unwanted dipole moment operators
in the decoupling procedure.  This enabled the treatment of superconductivity
on an equal footing with the Kondo effect. Moreover, Wu~\emph{et
  al.}~\cite{Wu-03} pointed out that the model of ultra-cold spin-3/2 fermions
with contact interaction enjoys a generic SP(4) symmetry, which lead to
further applications of the symplectic symmetry in the context of ultra-cold
fermionic gases~\cite{Lecheminant-05}.

In this paper we will combine these aspects and generalize the VBS state to
symplectic symmetry. We derive an exact parent Hamiltonian and argue that
there exists a finite gap in the excitation spectrum. We then use the
representation of the VBS state in terms of a matrix product state to
calculate the static correlation functions and the expectation values of
various string operators. Finally we discuss the VBS state and possible parent
Hamiltonians on higher-dimensional lattices.

\section{Symplectic symmetry}
One of the key features of the group SU(2) is that two spins of arbitrary
length $S$ can always combine into a singlet, which is an essential condition
for a proper description of frustrated antiferromagnetism. The analog
statement is not true for spins transforming under SU($n$) with $n\ge 3$,
where one has to deal with the tensor product of a representation and its
complex conjugate one in order to form a singlet. In the language of
antiferromagnetism this requires a bipartite lattice structure where one can
place spins transforming under one representation of SU($n$) on one sublattice
and the complex conjugated spins on the other sublattice.  One way to overcome
this problem~\cite{ReadSachdev91} is the generalization of SU(2) spins to
spins transforming under the symplectic group SP($n$), for which the formation
of a singlet from two spins is always possible.

The symplectic group SP($n$) is the set of all unitary $n\times
n$-matrices $U$ such that~\cite{Hamermesh89,Cornwell84vol2}
\begin{equation}
U^\mathrm{t}IU=I,
\label{eq:SP4def}
\end{equation}
where $^\mathrm{t}$ denotes the transposed matrix and 
\begin{equation}
I=\left(\begin{array}{ccccc}
0&1&\cdots&0&0\\-1&0&\cdots&0&0\\\vdots&\vdots&\ddots&\vdots&\vdots
\\0&0&\cdots&0&1\\0&0&\cdots&-1&0
\end{array}\right)\!.
\end{equation}
As the matrix $I$ is built up from blocks of $2\times 2$ matrices, $n$ has to
be even.  The generators of SP($n$), which we denote by $A^a$,
$a=1,\ldots,n(n+1)/2$, have to satisfy
\begin{equation}
\bigl(A^a\bigr)^\mathrm{t}I+IA^a=0.
\label{eq:sp4algebradef}
\end{equation}
The elements in the group are obtained by $U=\exp\left(\ii\,\sum_{a}\theta_a
  A^a\right)$ with real parameters $\theta_a$. The matrices $A^a$ play the
same role as the Pauli matrices for SU(2) and equal them in the case $n=2$.
Hence there exists an isomorphism between SP(2) and SU(2); in particular the
representations of SP(2) equal those of SU(2). An explicit representation of
the matrices $A^a$ for SP(4) is given in App.~\ref{sec:app1}. The irreducible
representations of SP($n$) can be labeled~\cite{Hamermesh89} by
$(\lambda_1\ldots\lambda_{n/2})$, where the non-negative integers $\lambda_i$
have to satisfy $\lambda_1\ge\lambda_2\ge\ldots\ge\lambda_{n/2}$.  Explicit
formulas for the dimensions of the irreducible representations, the
eigenvalues of the quadratic Casimir operator, and results on the
decomposition of tensor products of irreducible representations are stated in
App.~\ref{sec:app2}.  In Tab.~\ref{tab:reps} we have tabulated these
properties for those irreducible representations which we will use to
construct the VBS chain below.
\begin{table}[t]
\centering
\begin{tabular}{ccccc}
irreducible & \phantom{00} & dimension & \phantom{00} 
& eigenvalue \\[1mm]
representation & \phantom{00} & & \phantom{00} & of $\bs{J}^2$\\[1mm]
\hline
$(00\ldots 0)$ & \phantom{00} & $1$ & \phantom{00} & $0$\\[1mm]
$(10\ldots 0)$ & \phantom{00} & $n$ & \phantom{00} & $\frac{n+1}{4}$\\[1mm]
$(110\ldots 0)$ & \phantom{00} & $\frac{1}{2}(n-2)(n+1)$ & 
\phantom{00} & $\frac{n}{2}$\\[1mm]
$(20\ldots 0)$ & \phantom{00} & $\frac{n}{2}(n+1)$ & 
\phantom{00} & $\frac{n+2}{2}$\\[1mm]
$(220\ldots 0)$ & \phantom{00} & $\frac{n}{12}(n-2)(n-1)(n+3)$ & 
\phantom{00} & $n+1$\\[1mm]
$(310\ldots 0)$ & \phantom{00} & $\frac{n}{8}(n-2)(n+1)(n+3)$ & 
\phantom{00} & $n+2$\\[1mm]
$(40\ldots 0)$ & \phantom{00} & $\frac{n}{24}(n+1)(n+2)(n+3)$ & 
\phantom{00} & $n+4$
\end{tabular}
\caption{Simplest irreducible representations of SP($n$), their dimensions, 
  and the eigenvalues of the quadratic Casimir operator $\bs{J}^2$. We note 
  that the representations $(110\ldots 0)$, $(220\ldots 0)$, and 
  $(310\ldots 0)$ do not exist for SP(2) $\cong$ SU(2). In this case the 
  remaining representations are the singlet $\bs{0}$, the spinor 
  representation $\bs{\tfrac{1}{2}}$, the triplet $\bs{1}$, and the spin-2 
  representation $\bs{2}$.}
\label{tab:reps}
\end{table}

As a side note we mention that the symplectic group SP($n$) naturally arises
in Hamiltonian mechanics~\cite{Arnold89}. The $n$-dimensional phase space $M$
contains the generalized coordinates $q_1,\ldots,q_{n/2}$ and their conjugated
momenta $p_1,\ldots,p_{n/2}$, which implies that $n$ has to be even. The
Hamiltonian $H:M\rightarrow\mathbb{R}$ induces the time evolution via its
vector field. The phase space is equipped with a skew-scalar product on its
cotangent bundle $T^*M$, i.e., a bilinear map $\langle.,.\rangle:T^*M\times
T^*M\rightarrow\mathbb{R}$ which satisfies $\langle x,y\rangle=-\langle
y,x\rangle$. This skew-scalar product defines a volume element on the phase
space. The symplectic group is now the set of all linear transformation under
which this skew-scalar product is invariant. In particular, the time evolution
generated by the Hamiltonian is a symplectic transformation, which implies for
example Liouville's theorem. 

\section{Dimer chain}
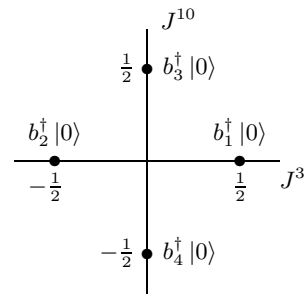
\begin{figure}[t]
\centering
\begin{picture}(110,120)
\thinlines
\put(0,50){\line(1,0){100}}
\put(50,0){\line(0,1){100}}
\multiput(15,50)(70,0){2}{\circle*{4}}
\multiput(50,15)(0,70){2}{\circle*{4}}
\put(100,40){$J^3$}
\put(55,100){$J^{10}$}
\put(32,13){$-\tfrac{1}{2}$}
\put(32,83){$\phantom{-}\tfrac{1}{2}$}
\put(5,38){$-\tfrac{1}{2}$}
\put(75,38){$\phantom{-}\tfrac{1}{2}$}
\put(75,58){$b_1^\dagger\ket{0}$}
\put(5,58){$b_2^\dagger\ket{0}$}
\put(56,83){$b_3^\dagger\ket{0}$}
\put(56,13){$b_4^\dagger\ket{0}$}
\end{picture}
\caption{Weight diagram of the fundamental representation of SP(4).  The 
  states are labeled using the bosonic creation operators introduced in 
  \eqref{eq:bosonbasis}. $J^3$ and $J^{10}$ denote the diagonal spin
  operators, their eigenvalues are easily obtained using 
  \eqref{eq:SP4operators}.}
\label{fig:fundamentl-weight}
\end{figure}

As a warmup exercise we first construct the SP($n$) generalization of
the Majumdar-Ghosh model~\cite{MajumdarGhosh69jmp1}.  Let
us consider a chain with $N$ lattice sites and periodic boundary conditions,
where we assume $N$ to be even. On each lattice site we place an SP($n$) spin
transforming under the fundamental, $n$-dimensional representation $(10\ldots
0)$. A basis at each lattice site $i$ may be written in terms of bosonic
creation and annihilation operators $b_{\sigma,i}^\dagger$ and $b_{\sigma,i}$
as~\cite{Holman69}
\begin{equation}
\ket{\sigma}_i=b_{\sigma,i}^\dagger\ket{0}_i,\quad \sigma=1,\ldots,n,
\label{eq:bosonbasis}
\end{equation}  
where $\ket{0}_i$ denotes the vacuum at site $i$.  The weight diagram of the
fundamental representation of SP(4) is shown in
Fig.~\ref{fig:fundamentl-weight}. 
The action of the SP($n$) spin operators
$J^a$ on these basis states is given by
\begin{equation}
\bs{J}_i=\frac{1}{2}\,\sum_{\sigma,\sigma'=1}^n\,
b_{\sigma,i}^{\dagger} \bs{A}_{\sigma\sigma'}^{\phantom{\dagger}}
b_{\sigma',i}^{\phantom{\dagger}}, 
\label{eq:s}
\end{equation}
where we have introduced the vector notation
$\bs{J}=(J^1,\ldots,J^{n(n+1)/2})^\mathrm{t}$. The eigenvalue of the quadratic
Casimir operator on each lattice site equals $\bs{J}^2_i=(n+1)/4$.  Using the
explicit expressions for the generators of SP(4) given in App.~\ref{sec:app1}
one finds for example
\begin{equation}
\begin{split}
&J^1_i=\frac{1}{2}\bigl(b_{1,i}^\dagger b_{2,i}^{\phantom{\dagger}}+
b_{2,i}^\dagger b_{1,i}^{\phantom{\dagger}}\bigr),\quad
J^2_i=\frac{\ii}{2}\bigl(b_{2,i}^\dagger b_{1,i}^{\phantom{\dagger}}-
b_{1,i}^\dagger b_{2,i}^{\phantom{\dagger}}\bigr),\\
&J^3_i=\frac{1}{2}\bigl(b_{1,i}^\dagger b_{1,i}^{\phantom{\dagger}}-
b_{2,i}^\dagger b_{2,i}^{\phantom{\dagger}}\bigr),\\
&J^4_i=\frac{1}{\sqrt{8}}\bigl(b_{1,i}^\dagger b_{4,i}^{\phantom{\dagger}}+
b_{2,i}^\dagger b_{3,i}^{\phantom{\dagger}}+
b_{3,i}^\dagger b_{2,i}^{\phantom{\dagger}}+
b_{4,i}^\dagger b_{1,i}^{\phantom{\dagger}}\bigr),\\
&J^5_i=\frac{\ii}{\sqrt{8}}\bigl(b_{4,i}^\dagger b_{1,i}^{\phantom{\dagger}}-
b_{3,i}^\dagger b_{2,i}^{\phantom{\dagger}}+
b_{2,i}^\dagger b_{3,i}^{\phantom{\dagger}}-
b_{1,i}^\dagger b_{4,i}^{\phantom{\dagger}}\bigr),\\
&J^6_i=\frac{1}{\sqrt{8}}\bigl(b_{1,i}^\dagger b_{3,i}^{\phantom{\dagger}}-
b_{2,i}^\dagger b_{4,i}^{\phantom{\dagger}}+
b_{3,i}^\dagger b_{1,i}^{\phantom{\dagger}}-
b_{4,i}^\dagger b_{2,i}^{\phantom{\dagger}}\bigr),\\
&J^7_i=\frac{\ii}{\sqrt{8}}\bigl(b_{1,i}^\dagger b_{3,i}^{\phantom{\dagger}}+
b_{2,i}^\dagger b_{4,i}^{\phantom{\dagger}}-
b_{3,i}^\dagger b_{1,i}^{\phantom{\dagger}}-
b_{4,i}^\dagger b_{2,i}^{\phantom{\dagger}}\bigr),\\
&J^8_i=\frac{1}{2}\bigl(b_{3,i}^\dagger b_{4,i}^{\phantom{\dagger}}+
b_{4,i}^\dagger b_{3,i}^{\phantom{\dagger}}\bigr),\quad
J^9_i=\frac{\ii}{2}\bigl(b_{4,i}^\dagger b_{3,i}^{\phantom{\dagger}}-
b_{3,i}^\dagger b_{4,i}^{\phantom{\dagger}}\bigr),\\
&J^{10}_i=\frac{1}{2}\bigl(b_{3,i}^\dagger b_{3,i}^{\phantom{\dagger}}-
b_{4,i}^\dagger b_{4,i}^{\phantom{\dagger}}\bigr).
\end{split}
\label{eq:SP4operators}
\end{equation}
We stress that the operators $J^1$, $J^2$,
and $J^3$ as well as $J^8$, $J^9$, and $J^{10}$ span two copies of the usual
spin algebra su(2).

On this SP($n$) chain we consider the two linearly independent states
represented by
\begin{equation}
\begin{array}{lc}
\ket{\hbox{\begin{picture}(116,8)(-4,-3)
\multiput(6,0)(14,0){8}{\circle{4}}
\thicklines
\multiput(8,0)(28,0){4}{\line(1,0){10}}
\end{picture}}} 
&\quad\textrm{``odd''}\\[2mm]
\ket{\hbox{\begin{picture}(116,8)(-4,-3)
\multiput(6,0)(14,0){8}{\circle{4}}
\thicklines
\multiput(22,0)(28,0){3}{\line(1,0){10}}
\put(1,0){\line(1,0){3}}
\put(106,0){\line(1,0){3}}
\end{picture}}} 
&\quad\textrm{``even''}
\end{array}
\label{eq:dimerstates}
\end{equation}
where the symbol 
\begin{picture}(26,5)(0,-3)
\put(6,0){\circle{4}}
\put(20,0){\circle{4}}
\thicklines
\put(8,0){\line(1,0){10}}
\end{picture}
stands for an SP($n$) singlet or dimer formed by the spins on two neighboring
lattice sites. In the state labeled as ``odd'' the SP($n$) singlets are
located on the bonds $(2i-1,2i)$, whereas in the state labeled as ``even'' the
SP($n$) singlets are located on the bonds $(2i,2i+1)$. In the ``even'' state
the right- and left-most spins also form an SP($n$) singlet due to the assumed
periodic boundary conditions.  In order to construct a parent Hamiltonian,
i.e., a Hamiltonian which has the two states \eqref{eq:dimerstates} as its
unique ground states, we note that the total SP($n$) spin on each three
neighboring sites has to contain a singlet and thus transforms under the
fundamental representation $(10\ldots 0)$. Hence, for all lattice sites $i$
the operator $(\bs{J}_i+\bs{J}_{i+1}+\bs{J}_{i+2})^2-(n+1)/4$ annihilates the
dimer states \eqref{eq:dimerstates}, and by taking the sum over all lattice
sites we arrive at
\begin{equation}
H_{\text{dimer}} = 
\sum_{i=1}^N \left(\bs{J}_i \bs{J}_{i+1} +
\frac{1}{2}\bs{J}_i \bs{J}_{i+2} +\frac{n+1}{8}\right).
\label{eq:dimerham}
\end{equation}
We have checked numerically for $n=4$ and $N=8$ that the model
\eqref{eq:dimerham} possesses exactly two zero-energy ground states. For $n=2$
one obtains the original Majumdar-Ghosh model~\cite{MajumdarGhosh69jmp1}.

\section{VBS chain}
In this section we construct the SP($n$) VBS state on a chain and derive the
corresponding parent Hamiltonian.  In the next sections we will then discuss
the excitations above the VBS state, its static correlation functions, and the
appearance of string order. 

Let us consider again a chain with $N$ lattice sites and periodic boundary
conditions, but now $N$ may be even or odd. At each lattice site we place two
copies of the fundamental representation $(10\ldots 0)$, i.e., we obtain the
tensor product (the decomposition of tensor products in irreducible
representations was derived in Refs.~\cite{Littelmann90,Leung93} and is
presented in App.~\ref{sec:app2})
\begin{equation}
(10\ldots 0)\otimes(10\ldots 0)=
(20\ldots 0)\oplus(110\ldots 0)\oplus(0\ldots 0).
\label{eq:tp2f}
\end{equation}
We note that for $n=2$ the representation $(110\ldots 0)$ does not
exist and we recover
$\bs{\tfrac{1}{2}}\otimes\bs{\tfrac{1}{2}}=\bs{1}\oplus\bs{0}$. In the
tensor product \eqref{eq:tp2f} we now project onto the adjoint,
$n(n+1)/2$-dimensional representation $(20\ldots 0)$. An explicit
basis for this representation can be constructed~\cite{Holman69} from
the bosonic basis of the fundamental representation
\eqref{eq:bosonbasis}.  For $n=4$ this basis will be stated explicitly
in Sec.~\ref{sec:corr}. With this procedure we construct a chain of
adjoint representations, which is the direct generalization of a
spin-1 chain for SU(2). If we consider the total SP($n$) spin of two
neighboring sites we find the decomposition
\begin{equation}
\begin{split}
&(20\ldots 0)\otimes(20\ldots 0)=
(40\ldots 0)\oplus(310\ldots 0)\\
&\qquad
\oplus(220\ldots 0)\oplus(20\ldots 0)\oplus(110\ldots 0)\oplus(0\ldots 0).
\end{split}
\label{eq:VBStp22}
\end{equation}
For $n=2$ the second, third, and fifth representation on the right-hand side
do not exist and \eqref{eq:VBStp22} simplifies to
$\bs{1}\otimes\bs{1}=\bs{2}\oplus\bs{1}\oplus\bs{0}$.

\begin{figure}[t]
\centering
\begin{picture}(210,51)
\multiput(10,35)(30,0){7}{\circle{4}}
\multiput(18,35)(30,0){7}{\circle{4}}
\multiput(14,35)(30,0){7}{\oval(20,12)}
\thicklines
\multiput(20,35)(30,0){6}{\line(1,0){18}}
\put(0,35){\line(1,0){8}}
\put(200,35){\line(1,0){8}}
\thinlines
\multiput(44,29)(0,-3){4}{\line(0,-1){2}}
\multiput(178,35)(0,-3){6}{\line(0,-1){2}}
\put(45,10){\makebox(0,0)[c]{\small projection onto $(20\ldots 0)$}}
\put(178,10){\makebox(0,0)[c]{\small singlet bond}}
\end{picture}
\caption{Graphical representation of the VBS state $\ket{\Psi_{\text{VBS}}}$, 
  the unique ground state of \eqref{eq:VBSham}. Each circle stands for a
  fundamental representation $(10\ldots 0)$, each line joining two circles for
  a singlet bond, and each oval for a lattice site on which we project onto 
  the adjoint representation $(20\ldots 0)$.}
\label{fig:VBSstate}
\end{figure}
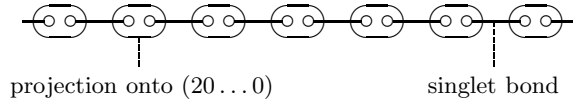

Starting with such a chain of adjoint representations, we can construct the
VBS state as follows: We form a singlet between one of the fundamental
representations $(10\ldots 0)$ on lattice site $i$ with one of the $(10\ldots
0)$'s on the neighboring site $i-1$ while we form another singlet with the
second representation $(10\ldots 0)$ on lattice site $i$ with one of the
$(10\ldots 0)$'s on the neighboring site $i+1$. We stress that the formation
of these singlets is imposed in addition to the already implemented projection
onto the adjoint representation at each lattice site. If we further impose
periodic boundary conditions this yields a unique VBS state
$\ket{\Psi_{\text{VBS}}}$, which is translationally invariant and can be
represented graphically as shown in Fig.~\ref{fig:VBSstate}.

The parent Hamiltonian for the VBS state is constructed by noting that on each
two neighboring sites in the VBS state we find one singlet and two uncoupled
fundamental representations. Hence, the total SP($n$) spin on two neighboring
sites is given by the tensor product $(10\ldots 0)\otimes(10\ldots 0)$ given
in \eqref{eq:tp2f}. If we construct an operator which is identical to zero on
\eqref{eq:tp2f} but takes strictly positive values on the complement of
\eqref{eq:tp2f} in $(20\ldots 0)\otimes(20\ldots 0)$, we will obtain the VBS
state as zero-energy ground state.  This operation is most easily implemented
using the quadratic Casimir operator $(\bs{J}_i+\bs{J}_{i+1})^2$ on the bond
$(i,i+1)$, which takes the values $(n+2)/2$, $n/2$, and $0$ on the
representations in \eqref{eq:tp2f} and $n+4$, $n+2$, and $n+1$ on the
remaining representations in \eqref{eq:VBStp22}, respectively.  Explicitly we
will use on each bond $(i,i+1)$:
\begin{equation}
\begin{split}
P_{i,i+1}&=\frac{1}{n}\,\frac{2}{5n^2+26n+32}\,
\biggl[\bigl(\bs{J}_{i}\!+\!\bs{J}_{i+1}\bigr)^2\biggr]\\[2mm]
&\;\times
\biggl[\bigl(\bs{J}_{i}\!+\!\bs{J}_{i+1}\bigr)^2-\frac{n+2}{2}\biggr]
\biggl[\bigl(\bs{J}_{i}\!+\!\bs{J}_{i+1}\bigr)^2-\frac{n}{2}\biggr].
\end{split}
\label{eq:VBSprojector}
\end{equation}
We stress that the operators $P_{i,i+1}$ are not simple projectors, as
$P_{i,i+1}$ takes different values on the subspaces $(220\ldots 0)$,
$(310\ldots 0)$, and $(40\ldots 0)$. We note that for $n=2$ the last factor in
\eqref{eq:VBSprojector} is not necessary as the corresponding representation
$(110\ldots 0)$ does not exist. The normalization of $P_{i,i+1}$ is chosen in
order to obtain a finite expectation value for the energy of each individual
bond in the limit $n\rightarrow \infty$. In this limit the operator
\eqref{eq:VBSprojector} becomes an orthogonal projector (up to the
multiplicative constant 1/10) onto the complement of \eqref{eq:tp2f} in
\eqref{eq:VBStp22}.  The parent Hamiltonian for the VBS state
$\ket{\Psi_{\text{VBS}}}$ is now obtained by $H=\sum_i P_{i,i+1}$ together
with $\bs{J}_i^2=(n+2)/2$:
\begin{equation}
\begin{split}
H=\frac{1}{n}\sum_{i=1}^N&\,\biggl[\bs{J}_{i}\,\bs{J}_{i+1}
+\frac{16n+40}{5n^2+26n+32}\bigl(\bs{J}_{i}\,\bs{J}_{i+1}\bigr)^2\\
+&\frac{16}{5n^2+26n+32}\bigl(\bs{J}_{i}\,\bs{J}_{i+1}\bigr)^3
+\frac{n^2+6n+8}{10n+32}\biggr].
\end{split}
\label{eq:VBSham}
\end{equation}
Here the operators $J_i^a$ live in the adjoined representation and can be
represented by $n(n+1)/2\times n(n+1)/2$-matrices.  As the operator
\eqref{eq:VBSprojector} takes strictly positive values on $(220\ldots 0)$,
$(310\ldots 0)$, and $(40\ldots 0)$, all states except the VBS state are
lifted to higher energies.  We have checked numerically for $n=4$ and $N=3$
that the VBS state is the unique ground state of \eqref{eq:VBSham}. A proof of
the uniqueness can be obtained by generalizing the proof of the uniqueness of
the ground state of the $q$-deformed VBS model~\cite{Kluemper-91}.  The
Hamiltonian contains cubic terms as we had to use three factors in the
operators \eqref{eq:VBSprojector}. As explained above the third factor is
superfluous for $n=2$, omitting it yields the original AKLT
model~\cite{Affleck-87}. By keeping the third factor, however, we obtain an
alternative parent Hamiltonian for the spin-1 VBS state.

The VBS construction described above can also be done for a chain with open
boundary conditions. In this case we are left with one uncoupled fundamental
representation at each end of the chain and we hence find $n^2$ linearly
independent VBS states.  The parent Hamiltonian for these states is given by
\eqref{eq:VBSham} with the summation restricted to $1\le i\le N-1$.

\section{Excitations and energy gap}
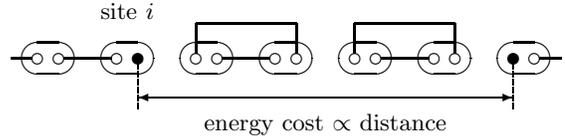
\begin{figure}[t]
\centering
\begin{picture}(210,65)
\put(34,50){site $i$}
\multiput(10,35)(30,0){7}{\circle{4}}
\multiput(18,35)(30,0){7}{\circle{4}}
\multiput(14,35)(30,0){7}{\oval(20,12)}
\multiput(48,35)(142,0){2}{\circle*{4}}
\thicklines
\put(20,35){\line(1,0){18}}
\multiput(80,35)(60,0){2}{\line(1,0){18}}
\multiput(70,48)(60,0){2}{\line(1,0){38}}
\multiput(70,37)(60,0){2}{\line(0,1){11}}
\multiput(108,37)(60,0){2}{\line(0,1){11}}
\put(0,35){\line(1,0){8}}
\put(200,35){\line(1,0){8}}
\thinlines
\multiput(48,33)(0,-3){6}{\line(0,-1){2}}
\multiput(190,33)(0,-3){6}{\line(0,-1){2}}
\put(119,20){\vector(-1,0){71}}
\put(119,20){\vector(1,0){71}}
\put(119,10){\makebox(0,0)[c]{\small energy cost $\propto$ distance}}
\end{picture}
\caption{If two spinons (represented by the full circles) move apart
  from each other, the region between them will not have the same structure as
  in the ground state (we have sketched one way of how the SP($n$) spins may
  rearrange into singlets). This causes an energy cost proportional to their
  distance and results in spinon confinement.}
\label{fig:confinement}
\end{figure}

The Hamiltonian \eqref{eq:VBSham} was constructed to be the exact
parent Hamiltonian for the VBS state $\ket{\Psi_{\text{VBS}}}$.
Although its ground state is known in all detail it is much harder to
get results on the excitations above it. The simplest operation on the
state $\ket{\Psi_{\text{VBS}}}$ one can imagine is to break one of the
singlets, say the singlet on the bond $(i,i+1)$. Doing so we find two
uncoupled SP($n$) spins each transforming under the fundamental
representation, which we will call spinons in the following. The
resulting state is clearly not an eigenstate of \eqref{eq:VBSham}.
Nevertheless, the spinons are useful to perform the following
Gedankenexperiment: Let us pin the first spinon at
site $i$ and move the other spinon to the right (see
Fig.~\ref{fig:confinement}).  The region between them has now a
different structure than the ground state and is not annihilated by
\eqref{eq:VBSham}.  As the energy cost grows linearly with the
distance, the spinons are subject to a linear confinement potential
and hence can only appear in bound states.  The relative motion of the
spinons will be described by a non-harmonic oscillator whose
zero-point energy yields a finite gap for the creation of
spinon-spinon bound states. This is consistent with the picture that
the origin of the Haldane gap is a confinement force between
spinons~\cite{GRS07,affleck:confinement}.  A similar argumentation was
applied by Greiter~\cite{Greiter02prb1} to the excitations of the
two-leg $t$-$J$ ladder. Although this Gedankenexperiment suggests the
appearance of an energy gap, we stress that the spinon bound states
may not constitute good trial wave functions for the actual low-lying
excitations in the model.

A possible way to prove the existence of a gap above the ground state is
provided by the extension of results by Knabe~\cite{Knabe88} on a class of
SU(2) VBS Hamiltonians including the original AKLT model (details of the
derivation are given in App.~\ref{sec:app3}). Let us consider a Hamiltonian of
the form
\begin{equation}
H=\sum_{i=1}^N\,P_{i,i+1}  
\label{eq:Knabeham}
\end{equation}
with periodic boundary conditions. We assume that $0\le P_{i,i+1}\le 1$ as
well as the existence of at least one zero-energy ground state of
\eqref{eq:Knabeham}.  The idea is to establish the inequality
\begin{equation}
H^2\ge \epsilon H,\quad \epsilon>0,
\label{eq:Knabecondition}
\end{equation}
which implies that the lowest non-vanishing eigenvalue of $H$ is larger than
$\epsilon$. As we show in App.~\ref{sec:app3}, Eq.~\eqref{eq:Knabecondition}
can be derived if the same model on a chain with $m+1$ sites and open boundary
conditions satisfies
\begin{equation}
h_{i,m}^2\ge \epsilon_m h_{i,m}, \quad \epsilon_m>\frac{1}{m},
\label{eq:Knabeconditionh}
\end{equation}
where $m\ge 2$ and 
\begin{equation}
h_{i,m}=\sum_{k=i}^{i+m-1}\,P_{k,k+1}.
\label{eq:Knabesmallham}
\end{equation}
Hence the proof is finished if we can show that \eqref{eq:Knabeconditionh} is
satisfied for a suitable integer $m$. This was achieved in Ref.~\cite{Knabe88}
for SU(2) VBS chains with spins $S=1,3/2,2$, and $5/2$ on each lattice site.
Unfortunately, exact diagonalization of the SP(4) model \eqref{eq:VBSham} with
open boundary conditions for $m=2$ suggests that in order to establish the
inequality $\epsilon_m>1/m$ one has to study chains with at least ten lattice
sites.

\section{Static correlation functions}
\label{sec:corr}
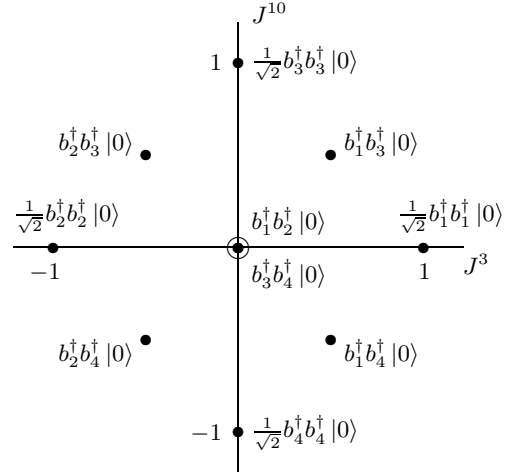
\begin{figure}[t]
\centering
\begin{picture}(180,190)
\thinlines
\put(0,85){\line(1,0){170}}
\put(85,0){\line(0,1){170}}
\put(85,85){\circle{8}}
\multiput(15,85)(70,0){3}{\circle*{4}}
\multiput(85,15)(0,70){3}{\circle*{4}}
\multiput(50,120)(70,0){2}{\circle*{4}}
\multiput(50,50)(70,0){2}{\circle*{4}}
\put(170,75){$J^3$}
\put(90,170){$J^{10}$}
\put(67,12){$-1$}
\put(67,152){$\phantom{-}1$}
\put(6,73){$-1$}
\put(146,73){$\phantom{-}1$}
\put(145,95){$\tfrac{1}{\sqrt{2}}b_1^\dagger b_1^\dagger\ket{0}$}
\put(0,95){$\tfrac{1}{\sqrt{2}}b_2^\dagger b_2^\dagger\ket{0}$}
\put(90,153){$\tfrac{1}{\sqrt{2}}b_3^\dagger b_3^\dagger\ket{0}$}
\put(90,13){$\tfrac{1}{\sqrt{2}}b_4^\dagger b_4^\dagger\ket{0}$}
\put(90,92){$b_1^\dagger b_2^\dagger\ket{0}$}
\put(90,72){$b_3^\dagger b_4^\dagger\ket{0}$}
\put(125,122){$b_1^\dagger b_3^\dagger\ket{0}$}
\put(125,42){$b_1^\dagger b_4^\dagger\ket{0}$}
\put(17,122){$b_2^\dagger b_3^\dagger\ket{0}$}
\put(17,42){$b_2^\dagger b_4^\dagger\ket{0}$}
\end{picture}
\caption{Weight diagram of the adjoint representation of SP(4).
  The state with $J^3=J^{10}=0$ is doubly degenerate. The states are labeled
  using the bosonic creation operators as in \eqref{eq:bosonicbasis10}.}
\label{fig:adjoint-weight}
\end{figure}

The VBS state $\ket{\Psi_{\text{VBS}}}$ can be written as a matrix product
state. We will restrict ourselves to the case $n=4$ in the following. A
suitable basis for the adjoint representation of SP(4) at lattice site $i$ can
be obtained from the bosonic basis of the fundamental representation
introduced above as~\cite{Holman69}
\begin{equation}
  \begin{split}
  &\frac{1}{\sqrt{2}}b_{1,i}^\dagger b_{1,i}^\dagger\ket{0}_i,\;
  \frac{1}{\sqrt{2}}b_{2,i}^\dagger b_{2,i}^\dagger\ket{0}_i,\;
  b_{1,i}^\dagger b_{2,i}^\dagger\ket{0}_i,\\
  &b_{1,i}^\dagger b_{3,i}^\dagger\ket{0}_i,\;
  b_{1,i}^\dagger b_{4,i}^\dagger\ket{0}_i,\;
  b_{2,i}^\dagger b_{3,i}^\dagger\ket{0}_i,\;
  b_{2,i}^\dagger b_{4,i}^\dagger\ket{0}_i,\\
  &\frac{1}{\sqrt{2}}b_{3,i}^\dagger b_{3,i}^\dagger\ket{0}_i,\;
  \frac{1}{\sqrt{2}}b_{4,i}^\dagger b_{4,i}^\dagger\ket{0}_i,\;
  b_{3,i}^\dagger b_{4,i}^\dagger\ket{0}_i.
  \end{split}
\label{eq:bosonicbasis10}
\end{equation}
We have illustrated these basis states in the weight diagram of the adjoint
representation shown in Fig.~\ref{fig:adjoint-weight}.

In order to derive the matrix product representation we first rewrite the
singlet on the bond $(i,i+1)$ as
\begin{equation}
\begin{split}
&b^\dagger_{1,i} b^\dagger_{2,i+1} - 
b^\dagger_{2,i} b^\dagger_{1,i+1} +
b^\dagger_{3,i} b^\dagger_{4,i+1} - 
b^\dagger_{4,i} b^\dagger_{3,i+1} \\[2mm]
&\quad
=\Bigl(b^\dagger_{1,i},b^\dagger_{2,i},b^\dagger_{3,i},b^\dagger_{4,i}\Bigr)
\left(\!\begin{array}{c} \phantom{-}b^\dagger_{2,i+1}\\
-b^\dagger_{1,i+1}\\ \phantom{-}b^\dagger_{4,i+1}\\
-b^\dagger_{3,i+1}\end{array}\!\right).
\end{split}
\label{eq:VBSsinglet}
\end{equation}
Second, at each lattice site $i$ we use the outer product to combine the two
vectors originating from the rewriting \eqref{eq:VBSsinglet} on the bonds
$(i-1,i)$ and $(i,i+1)$ into a matrix
\begin{equation}
\begin{split}
M_i&=\,\left(\!\begin{array}{c} \phantom{-}b^\dagger_{2,i}\\
-b^\dagger_{1,i}\\ \phantom{-}b^\dagger_{4,i}\\
-b^\dagger_{3,i}\end{array}\!\right)
\Bigl(b^\dagger_{1,i},b^\dagger_{2,i},b^\dagger_{3,i},
b^\dagger_{4,i}\Bigr)\ket{0}_i\\[2mm]
&=\left(\begin{array}{cccc}
\phantom{-}b_{1,i}^\dagger b_{2,i}^\dagger & 
\phantom{-}b_{2,i}^\dagger b_{2,i}^\dagger &
\phantom{-}b_{2,i}^\dagger b_{3,i}^\dagger & 
\phantom{-}b_{2,i}^\dagger b_{4,i}^\dagger\\
-b_{1,i}^\dagger b_{1,i}^\dagger & -b_{1,i}^\dagger b_{2,i}^\dagger &
-b_{1,i}^\dagger b_{3,i}^\dagger & -b_{1,i}^\dagger b_{4,i}^\dagger\\
\phantom{-}b_{1,i}^\dagger b_{4,i}^\dagger & 
\phantom{-}b_{2,i}^\dagger b_{4,i}^\dagger &
\phantom{-}b_{3,i}^\dagger b_{4,i}^\dagger & 
\phantom{-}b_{4,i}^\dagger b_{4,i}^\dagger\\
-b_{1,i}^\dagger b_{3,i}^\dagger & -b_{2,i}^\dagger b_{3,i}^\dagger &
-b_{3,i}^\dagger b_{3,i}^\dagger & -b_{3,i}^\dagger b_{4,i}^\dagger
\end{array}
\right)\ket{0}_i.
\end{split}
\label{eq:VBSmatrix}
\end{equation}
Assuming periodic boundary conditions the VBS state can then be written as the
trace of the matrix product
\begin{equation}
\label{eq:mps}
\ket{\Psi_{\text{VBS}}}=\text{tr}\Biggl( \prod_{i=1}^N M_i \Biggr).
\end{equation}
Starting from this representation the static correlation functions in the
SP(4) VBS state can be calculated by applying the method introduced by
Kl\"umper~\emph{et al.} for the analysis of the $q$-deformed
model~\cite{Kluemper-91}. As the first step we calculate the norm of the VBS
state. This is done by introducing the complex conjugated matrix $\tilde{M}$
according to $\tilde{M}_{\sigma\sigma'}=M^*_{\sigma\sigma'}$, i.e., by simply
taking the complex conjugate of each matrix element in \eqref{eq:VBSmatrix}
without transposing the matrix. We then define the $16\times 16$ transfer
matrix $R$ at any lattice site as
\begin{equation}
\label{eq:Rmatrix}
R_{\alpha\beta}=R_{(\sigma\tau),(\sigma'\tau')}=
\tilde{M}_{\sigma\sigma'}\,M_{\tau\tau'},
\end{equation}
where we order the indices as $\alpha,\beta=1,\ldots,16 \leftrightarrow
(11),(12),\ldots,(44)$. The norm of the VBS state is now given by
\begin{equation}
\label{eq:VBSnorm}
\bra{\Psi_{\text{VBS}}}\Psi_{\text{VBS}}\rangle=
\text{tr}\Bigl(R^N\Bigr)=5^N+10\,(-1)^N+5,
\end{equation}
where we have evaluated the trace by diagonalization of $R$.  In the second
step we calculate the expectation value
$\bra{\Psi_{\text{VBS}}}J^3_1\,J^3_j\ket{\Psi_{\text{VBS}}}$. We introduce the
transfer-matrix representation of the spin operators $J^3$ by
\begin{equation}
\hat{J}_{\alpha\beta}=\hat{J}_{(\sigma\tau),(\sigma'\tau')}=
\tilde{M}_{\sigma\sigma'}\,J^3\,M_{\tau\tau'}.
\end{equation}
Here the operator $J^3$ acts on the elements of $M$ as 
\begin{equation}
J^3b_1^\dagger=\frac{1}{2}b_1^\dagger,\quad
J^3b_2^\dagger=-\frac{1}{2}b_2^\dagger,\quad
J^3b_3^\dagger=J^3b_4^\dagger=0,
\end{equation}
which implies for example $J^3b_1^\dagger b_2^\dagger=0$.
This yields 
\begin{equation}
\bra{\Psi_{\text{VBS}}}J^3_1\,J^3_j\ket{\Psi_{\text{VBS}}}=
\text{tr}\Bigl(\hat{J_1}\,R^{j-2}\,\hat{J_j}\,R^{N-j}\Bigr),
\end{equation}
which is easily evaluated by diagonalization of $R$.  As the state
$\ket{\Psi_{\text{VBS}}}$ enjoys full SP(4) symmetry we arrive at
\begin{equation}
\label{eq:VBScorr}
\begin{split}
&\Big\langle J^a_1\,J^b_j\Big\rangle\equiv
\frac{\bra{\Psi_{\text{VBS}}}J^a_1\,J^b_j\ket{\Psi_{\text{VBS}}}}
{\bra{\Psi_{\text{VBS}}}\Psi_{\text{VBS}}\rangle}\\
&\quad=-\delta_{ab}\,
(-1)^j\,\frac{9}{10}\,
\frac{5^{-j+1}+\frac{1}{5}\frac{(-1)^{N}}{5^{N-j}}+
\frac{1}{3}\frac{(-1)^{N}+1}{5^{N-1}}}
{1+\frac{1}{5^{N-1}}\bigl(2 (-1)^N+1\bigr)}.
\end{split}
\end{equation}
In the general case of SP($n$) the same steps yield in the thermodynamic limit
$N\rightarrow\infty$
\begin{equation}
\Big\langle J^a_1\,J^b_j\Big\rangle\propto
\frac{\delta_{ab}}{(n+1)^{j-1}}\sim e^{-j/\xi}.
\end{equation}
Here the correlation length is given by $\xi=1/\ln(n+1)$ and vanishes in the
limit $n\rightarrow\infty$.  We also recover the known result for the AKLT
chain~\cite{Affleck-87}.

\section{String order}
It is well known~\cite{denNijsRommelse89} that there exists a hidden nonlocal
topological order or string order in the AKLT model.  In fact, this order is
found in the whole Haldane phase in the phase diagram of the general spin-1
chain. This string order was further associated with the breaking of a
$\mathbb{Z}_2\times\mathbb{Z}_2$-symmetry in the Haldane phase and the
appearance of a four-fold degenerate ground state on the open
chain~\cite{KennedyTasaki92prb}. We will find a similar behavior in the
SP($n$) VBS model.

In analogy to Refs.~\cite{denNijsRommelse89} we define the string operators
\begin{equation}
O_{1j}^{ab}=-J_1^a\,
\exp\left(\ii\pi\sum_{k=2}^{j-1}\sum_{c}J^c_k\right)\,J^b_j,
\label{eq:stringoperators}
\end{equation}
where the second sum is over all $c$ for which $J^c_k$ is diagonal and $J_1^a$
and $J_j^b$ have to be diagonal as well.  In the SP(4) model the summation is
over $c=3,10$ and we have $a,b\in\{3,10\}$.  Using the transfer-matrix
technique we obtain in the thermodynamic limit $N\rightarrow\infty$
\begin{equation}
\begin{split}
&\Big\langle O_{1j}^{33}\Big\rangle=\Big\langle O_{1j}^{10,10}\Big\rangle=
\frac{9}{100}\left(1+\frac{25}{5^j}\right),\\
&\Big\langle O_{1j}^{3,10}\Big\rangle=\Big\langle O_{1j}^{10,3}\Big\rangle=
\frac{9}{100}\left(1-\frac{25}{5^j}\right),
\end{split}
\label{eq:stringorder}
\end{equation}
which remain finite for arbitrary large values of $j$. In particular, the sum
over all four expectation values \eqref{eq:stringorder} is independent of $j$.
In analogy to the original AKLT model we expect this hidden string order as
well as the 16-fold degeneracy of the ground state of a chain with open
boundary conditions to be a consequence of the breaking of a discrete symmetry
($\mathbb{Z}_4\times\mathbb{Z}_4$).

We have also calculated the expectation values of the nine string operators
\eqref{eq:stringoperators} in the SP(6) model. Together with
\eqref{eq:stringorder} and the result~\cite{denNijsRommelse89} for SU(2) this
leads us to the conjecture for general $n$:
\begin{equation}
\begin{split}
&\Big\langle O_{1j}^{aa}\Big\rangle=
\frac{\bigl(\frac{n}{2}+1\bigr)^2}{\bigl(\frac{n}{2}(n+1)\bigr)^2}
\left(1+\frac{\frac{n}{2}-1}{(n+1)^{j-2}}\right),\\
&\Big\langle O_{1j}^{ab}\Big\rangle=
\frac{\bigl(\frac{n}{2}+1\bigr)^2}{\bigl(\frac{n}{2}(n+1)\bigr)^2}
\left(1-\frac{1}{(n+1)^{j-2}}\right),\quad a\neq b.
\end{split}
\label{eq:stringorderSPn}
\end{equation}
Although each of the expectation values \eqref{eq:stringorderSPn} vanishes in
the limit $n\rightarrow\infty$, the number of string operators increases and
one obtains
\begin{equation}
\sum_{a,b}\Big\langle O_{1j}^{ab}\Big\rangle=
\frac{1}{4}\left(\frac{n+2}{n+1}\right)^2
\rightarrow\frac{1}{4},\quad n\rightarrow\infty,
\label{eq:stringsum}
\end{equation}
where the sum is over all $a$ and $b$ for which $J^a$ and $J^b$ are diagonal.
We note that \eqref{eq:stringsum} can be written elegantly as a single string
operator by replacing $J^a_{1}$ and $J^b_{j}$ in \eqref{eq:stringoperators} by
the sum over all diagonal generators $\sum_c J^c_{1,j}$, respectively.

\pagebreak
\section{Two-dimensional VBS model}
\begin{figure}[t]
\centering
\begin{picture}(220,145)
\multiput(20,60)(68,0){3}{\circle{4}}
\multiput(28,60)(68,0){3}{\circle{4}}
\multiput(24,67)(68,0){3}{\circle{4}}
\multiput(24,63)(68,0){3}{\circle{18}}
\multiput(20,103)(68,0){3}{\circle{4}}
\multiput(28,103)(68,0){3}{\circle{4}}
\multiput(24,96)(68,0){3}{\circle{4}}
\multiput(24,100)(68,0){3}{\circle{18}}
\multiput(54,49)(68,0){2}{\circle{4}}
\multiput(62,49)(68,0){2}{\circle{4}}
\multiput(58,42)(68,0){2}{\circle{4}}
\multiput(58,47)(68,0){2}{\circle{18}}
\multiput(54,114)(68,0){2}{\circle{4}}
\multiput(62,114)(68,0){2}{\circle{4}}
\multiput(58,121)(68,0){2}{\circle{4}}
\multiput(58,117)(68,0){2}{\circle{18}}
\thicklines
\multiput(24,69)(68,0){3}{\line(0,1){25}}
\multiput(30,59)(68,0){2}{\line(5,-2){22}}
\multiput(64,50)(68,0){2}{\line(5,2){22}}
\multiput(58,40)(68,0){2}{\line(0,-1){10}}
\multiput(30,104)(68,0){2}{\line(5,2){22}}
\multiput(64,113)(68,0){2}{\line(5,-2){22}}
\multiput(58,123)(68,0){2}{\line(0,1){10}}
\put(18,59){\line(-5,-2){10}}
\put(166,104){\line(5,2){10}}
\put(18,104){\line(-5,2){10}}
\put(166,59){\line(5,-2){10}}
\thinlines
\multiput(24,52)(0,-3){8}{\line(0,-1){2}}
\multiput(160,81)(3,0){10}{\line(1,0){2}}
\put(2,20){\small projection}
\put(-4,10){\small onto $(30\ldots 0)$}
\put(195,85){\small singlet}
\put(198,75){\small bond}
\end{picture}
\caption{Graphical representation of the VBS state on a hexagonal lattice. 
  Each small circle represents a fundamental representation $(10\ldots 0)$,
  each line joining two circles for a singlet bond, and each large circle a
  lattice site on which we project onto the representation $(30\ldots 0)$.}
\label{fig:hexagonalVBSstate}
\end{figure}
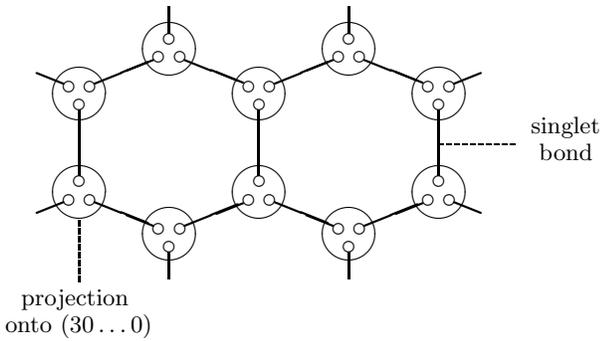
Finally we would like to discuss the VBS construction on higher-dimensional
lattices. The simplest example is provided by the honeycomb lattice (with
coordination number $z=3$) with representations $(30\ldots 0)$ on each lattice
site. The VBS state is obtained by placing three fundamental representations
on each lattice site and projecting onto the representation $(30\ldots 0)$ as
well as forming a singlet of each one of them with a fundamental
representation on a neighboring site (see Fig.~\ref{fig:hexagonalVBSstate}).
Hence, on each bond we obtain a singlet formed in this way and four uncoupled
fundamental representations.  The corresponding tensor product decomposes as
\begin{equation}
\begin{split} 
&(10\ldots 0)^{\otimes 4}=(40\ldots 0)\oplus 3\cdot(310\ldots 0)\\[1mm]
&\qquad\oplus 2\cdot(220\ldots 0)\oplus3\cdot(2110\ldots 0)
\oplus 6\cdot(20\ldots 0)\\[1mm]
&\qquad\oplus(11110\ldots 0)\oplus 6\cdot(110\ldots 0)\oplus 3\cdot(0\ldots 0).
\end{split}
\label{eq:hexagonaltp}
\end{equation}
Since this tensor product contains in general eight different irreducible
representations, the ``projection'' operator analog to \eqref{eq:VBSprojector}
and hence the Hamiltonian contains the bond operators
$\bigl(\bs{J}_i\,\bs{J}_j\bigr)^k$ with a power up to $k=8$.  For the cases
$n=2,4$, and $6$, however, some representations on the right-hand side of
\eqref{eq:hexagonaltp} do not exists and one obtains powers up to $k=3,6$, and
$7$, respectively. The explicit construction of the analog SU(2) model with
$S=3/2$ spins on the lattice sites can be found in Refs.~\cite{Affleck-87}. It
is clear from the arguments above that the VBS construction on lattices with
larger coordination number or in higher dimensions will lead to a parent
Hamiltonian which contains even higher powers of the bond operators
$\bigl(\bs{J}_i\,\bs{J}_j\bigr)^k$.

\section{Conclusions}
In conclusion, we have introduced a spin chain with symplectic
symmetry SP($n$) which shares all properties of the Haldane scenario
for integer-spin quantum antiferromagnets: (i) a unique ground state,
(ii) a finite gap in the energy spectrum above the ground state, and
(iii) ground-state correlation functions which fall off exponentially.
Furthermore we have shown that the ground state possesses string
order. We point out that in the limit $n\rightarrow\infty$ the string
order remains finite and the correlation length vanishes. The
application of the large-$n$ approach to the considered models might
be an interesting extension of this work.

\section*{Acknowledgments}
We would like to thank Piers Coleman, Fabian Essler, and Paul Fendley for
useful discussions.  We are especially grateful to Martin Greiter for numerous
discussions and sharing his expertise on VBS states with us.  This work was
supported by the Deutsche Akademie der Naturforscher Leopoldina under grant
BMBF-LPD 9901/8-145 (DS) and by a Ph.D. scholarship of the Cusanuswerk (SR).

\appendix
\section{Explicit representations for SP(4)}\label{sec:app1}
An explicit representation of the generators of SP(4) is provided by
\begin{displaymath}
\begin{split}
&A^1=\left(\begin{array}{cccc}
0&1&0&0\\1&0&0&0\\0&0&0&0\\0&0&0&0
\end{array}\right)\!\!,\;
A^2 =\left(\begin{array}{cccc}
0&-\text{i}&0&0\\\text{i}&0&0&0\\0&0&0&0\\0&0&0&0
\end{array}\right)\!\!,
\end{split}
\end{displaymath}
\begin{displaymath}
\begin{split}
&A^3 =\left(\begin{array}{cccc}
1&0&0&0\\0&-1&0&0\\0&0&0&0\\0&0&0&0
\end{array}\right)\!\!,\;
A^4=\frac{1}{\sqrt{2}}\left(\begin{array}{cccc}
0&0&0&1\\0&0&1&0\\0&1&0&0\\1&0&0&0
\end{array}\right)\!\!,
\end{split}
\end{displaymath}
\begin{displaymath}
\begin{split}
&A^5 =\frac{1}{\sqrt{2}}\left(\begin{array}{cccc}
0&0&0&-\text{i}\\0&0&\text{i}&0\\0&-\text{i}&0&0\\\text{i}&0&0&0
\end{array}\right)\!\!,\;
A^6 =\frac{1}{\sqrt{2}}\left(\begin{array}{cccc}
0&0&1&0\\0&0&0&-1\\1&0&0&0\\0&-1&0&0
\end{array}\right)\!\!,
\end{split}
\end{displaymath}
\begin{displaymath}
\begin{split}
&A^7=\frac{1}{\sqrt{2}}\left(\begin{array}{cccc}
0&0&\text{i}&0\\0&0&0&\text{i}\\-\text{i}&0&0&0\\0&-\text{i}&0&0
\end{array}\right)\!\!,\;
A^8 =\left(\begin{array}{cccc}
0&0&0&0\\0&0&0&0\\0&0&0&1\\0&0&1&0
\end{array}\right)\!\!,
\end{split}
\end{displaymath}
\begin{displaymath}
\begin{split}
&A^9 =\left(\begin{array}{cccc}
0&0&0&0\\0&0&0&0\\0&0&0&-\text{i}\\0&0&\text{i}&0
\end{array}\right)\!\!,\;
A^{10} =\left(\begin{array}{cccc}
0&0&0&0\\0&0&0&0\\0&0&1&0\\0&0&0&-1
\end{array}\right)\!\!.
\end{split}
\end{displaymath}
The normalization is chosen to be 
\begin{equation}
\mathrm{tr}\left(A^a A^b\right)=2\delta_{ab}.
\end{equation}  
The matrices $A^a$, $a=1,\ldots,10$, form a basis of sp(4), the Lie algebra of
SP(4). They satisfy the commutation relations
\begin{equation}
\comm{A^a}{A^b}=2f^{abc}A^c.
\end{equation}
The structure constants $f^{abc}$ are totally antisymmetric and obey Jacobi's
identity
\begin{equation}
f^{abc}f^{cde}+f^{bdc}f^{cae}+f^{dac}f^{cbe}=0.
\label{eq:appsu3-JI}
\end{equation} 
Explicitly, all 84 non-vanishing structure constants are obtained by
permutations of the indices from
\begin{equation}
\begin{split}
&f^{123}=f^{89,10}=\text{i},\\
&f^{156}=f^{345}=f^{45,10}=f^{478}=f^{568}=f^{579}=f^{67,10}
=\frac{\text{i}}{2},\\
&f^{147}=f^{246}=f^{257}=f^{367}=f^{469}=-\frac{\text{i}}{2}.
\end{split}
\end{equation}
sp(4) has rank two, the Cartan subalgebra is spanned by $A^3$ and $A^{10}$. We
note that sp(2)$\cong$ su(2) and sp(4)$\cong$ so(5). A possible matrix
representation of the spin operators in the adjoint representation is given by
$\bigl(J^a\bigr)_{bc}=f^{abc}$. However, we stress that these matrices are not
the representation matrices in the bosonic basis \eqref{eq:bosonicbasis10}.

\section{Some results on the representation theory of SP($n$)}
\label{sec:app2}
In this appendix we review some results on the representation theory of
SP($n$). First, the dimension of the irreducible representation
$(\lambda_1\ldots\lambda_{n/2})$ is given by the formula~\cite{Hamermesh89}
\begin{equation}
\begin{split}
&\dim\bigl[(\lambda_1\ldots\lambda_{n/2})\bigr]=
\prod_{i=1}^{n/2}\frac{\lambda_i+\frac{n}{2}-i+1}{\frac{n}{2}-i+1}\\
&\;\times\prod_{\substack{i,j=1\\i<j}}^{n/2}\frac{\lambda_i-\lambda_j+j-i}{j-i}
\frac{\lambda_i+\lambda_j+n+2-i-j}{n+2-i-j}.
\end{split}
\label{eq:appdim}
\end{equation}

Second, the eigenvalues of the quadratic Casimir operator $\bs{J}^2$ were
derived by Nwachuku and
Rashid~\cite{NwachukuRashid76} and read using our
conventions 
\begin{equation}
\bs{J}^2=\frac{1}{8}\sum_{\substack{i=-n/2\\i\neq 0}}^{n/2} \kappa_i^2
\frac{\kappa_i-\frac{n}{2}-1}{\kappa_i-\frac{n}{2}-\frac{1}{2}}
\prod_{\substack{j=-n/2\\j\neq 0,i}}^{n/2}
\left(1-\frac{1}{\kappa_i-\kappa_j}\right),
\end{equation}
where for $1\le i\le n/2$
\begin{equation}
\kappa_i=\frac{n}{2}+i+\lambda_{n/2+1-i},\quad
\kappa_{-i}=n-\kappa_i. 
\label{eq:appcas1} 
\end{equation}
The following special cases allow a closed expression:
\begin{eqnarray}
(\nu 0\ldots 0):& &\frac{\nu}{4}(n+\nu),\label{eq:appcas2}\\
(\underbrace{\nu\ldots\nu}_{k\;\text{terms}} 0\ldots 0):
& & \frac{\nu k}{4}(n+\nu-k+1).\label{eq:appcas3}
\end{eqnarray}
The relation to Refs.~\cite{NwachukuRashid76} is obtained by replacing
$n\rightarrow n/2$ and rescaling the Casimir operator by a factor of $1/8$.
For the simplest irreducible representations the formulas
\eqref{eq:appdim}--\eqref{eq:appcas3} easily yield the results stated in
Tab.~\ref{tab:reps}.

Finally, we make use of the following results on the decomposition of tensor
products into irreducible representations, which is in its general form due to
Littelmann~\cite{Littelmann90} and was specialized to the case we use here by
Leung~\cite{Leung93}:
\begin{equation}
\begin{split}
&(\mu_1\mu_2\ldots\mu_{n/2})\otimes(\nu 0\ldots0)\\[2mm]
&\quad =\sum_{\kappa_i}\bigoplus(\mu_1+\kappa_1-\kappa_n,
\mu_2+\kappa_2-\kappa_{n-1},\\[-2mm]
&\qquad\qquad\qquad\qquad,\ldots,\mu_{n/2}+\kappa_{n/2}-\kappa_{n/2+1}),
\end{split}
\end{equation}
where the sum is over all integers $\kappa_1,\ldots,\kappa_n$ subject to the
conditions:
\begin{displaymath}
\begin{split}
&\kappa_1+\ldots+\kappa_n=\nu,\\
&0\le\kappa_i\le\mu_{i-1}-\mu_i-\kappa_{n-i+2}+\kappa_{n-i+1},\\
&0\le\kappa_{n-j}\le\mu_{j+1}-\mu_{j+2},\\
&0\le\kappa_{n/2+1}\le\mu_{n/2}.
\end{split}
\end{displaymath}
where $i=2,3,\ldots,n/2$ and $j=0,1,\ldots,n/2-2$.

\section{Derivation of Eq. (\ref{eq:Knabecondition})}
\label{sec:app3}
In this appendix we will generalize results obtained by Knabe~\cite{Knabe88}
on the existence of a gap in SU(2) VBS chains with arbitrary spin. Similar
techniques were also used by Fannes \emph{et al.}~\cite{Fannes-91}.  The main
difference of our result as compared to Ref.~\cite{Knabe88} is that the
operators $P_{i,i+1}$ are not assumed to be simple projectors.

Let us start with \eqref{eq:Knabeham}. The assumption $P_{i,i+1}\le 1$ yields
$P^2_{i,i+1}\le P_{i,i+1}$, where inequalities between operators are
understood in the sense
\begin{equation}
\bra{\psi}P^2_{i,i+1}\ket{\psi}\le 
\bra{\psi}P_{i,i+1}\ket{\psi}\le
\bra{\psi}\psi\rangle
\end{equation}
for all states $\ket{\psi}$. In fact, the most useful results will be obtained
if the largest eigenvalue of $P_{i,i+1}$ equals one, which is obtained by
multiplication of \eqref{eq:VBSprojector} with a suitable constant. Using the
definitions \eqref{eq:Knabeham} and \eqref{eq:Knabesmallham} one easily finds
\begin{equation}
\begin{split}
&H^2= -\frac{1}{m-1}\sum_{i=1}^N\,P_{i,i+1}^2+
\frac{1}{m-1}\sum_{i=1}^N h_{i,m}^2\\
&+\sum_{\substack{i,j=1\\|i-j|>1}}^N\,P_{i,i+1}P_{j,j+1}-
\frac{1}{m-1}\sum_{i=1}^N
\sum_{\substack{k,l=i\\|k-l|>1}}^{i+m-1}\,P_{k,k+1}P_{l,l+1}
\end{split} 
\label{eq:Knabeapp}
\end{equation}
We can now use $P_{i,i+1}^2\le P_{i,i+1}$, which implies
$-\frac{1}{m-1}\sum_i\,P_{i,i+1}^2\ge -\frac{1}{m-1}H$, together with the fact
that each of the terms $P_{i,i+1}P_{j,j+1}$ appears more often in the third
sum than in the fourth sum. Therefore we get the inequality
\begin{equation}
H^2\ge\frac{1}{m-1}\sum_{i=1}^N h_{i,m}^2-\frac{1}{m-1}H.
\end{equation}
Finally we can use \eqref{eq:Knabeconditionh} as well as $\sum_i h_{i,m}=mH$
to obtain \eqref{eq:Knabecondition} with
$\epsilon=\frac{m}{m-1}\left(\epsilon_m-\frac{1}{m}\right)$.


\begin{thebibliography}{10}
  
\bibitem{Affleck-87} 
I. Affleck, T. Kennedy, E.~H. Lieb, and H. Tasaki, 
Phys. Rev. Lett.  {\bf 59}, 799 (1987);
Commun. Math. Phys. {\bf 115},  477  (1988).

\bibitem{Haldane83pl1}
F.~D.~M. Haldane, Phys. Lett.~A {\bf 93},  464  (1983);
Phys. Rev. Lett. {\bf 50},  1153  (1983);
I. Affleck, J.~Phys.: Condens. Matter {\bf 1}, 3047 (1989).

\bibitem{Arovas-88}
D.~P. Arovas, A. Auerbach, and F.~D.~M. Haldane, 
Phys. Rev. Lett. {\bf 60}, 531 (1988).

\bibitem{Auerbach94}
A. Auerbach, {\it Interacting electrons and quantum magnetism} (Springer, 
New York, 1994).

\bibitem{Laughlin83}
R. B. Laughlin, Phys. Rev. Lett. {\bf 50}, 1395 (1983);
F.~D.~M. Haldane, Phys. Rev. Lett. {\bf 51}, 605 (1983).

\bibitem{denNijsRommelse89}
M. den Nijs and K. Rommelse, Phys. Rev.~B {\bf 40},  4709  (1989);
S.~M. Girvin and D.~P. Arovas, Physica Scripta {\bf T27},  156  (1989).

\bibitem{KennedyTasaki92prb}
T. Kennedy and H. Tasaki, Phys. Rev.~B {\bf 45},  304  (1992);
Commun. Math. Phys. {\bf 147},  431  (1992);
M. Oshikawa, J.~Phys.: Condens. Matter {\bf 4},  7469  (1992).

\bibitem{Kluemper-91}
A. Kl\"umper, A. Schadschneider, and J. Zittartz, 
J.~Phys.~A: Math. Gen. {\bf  24},  L955  (1991);
Z.~Phys.~B {\bf 87},  281  (1992).

\bibitem{Affleck-91}
I. Affleck, D.~P. Arovas, J.~B. Marston, and D.~A. Rabson, Nucl. Phys.~B
{\bf 366}, 467 (1991);
D.~P. Arovas, Phys. Rev.~B {\bf 77},  104404  (2008);
H. Katsura, T. Hirano, and V.~E. Korepin, J.~Phys.~A: Math. Theor. {\bf 41},
  135304  (2008).

\bibitem{GRS07}
M. Greiter, S. Rachel, and D. Schuricht, 
Phys. Rev.~B {\bf 75},  060401(R)  (2007);
M. Greiter and S. Rachel, Phys. Rev.~B {\bf 75},  184441  (2007).

\bibitem{Tu-08}
H.-H. Tu, G.-M. Zhang, and T. Xiang, arXiv:0804.1685 [cond-mat.str-el]; 
arXiv:0806.1839 [cond-mat.str-el].

\bibitem{Fannes-91}
M. Fannes, B. Nachtergaele, and R.~F. Werner, 
J.~Phys.~A: Math. Gen. {\bf 24},  L185  (1991);
Commun. Math. Phys. {\bf 144},  443  (1992).

\bibitem{Kluemper-93}
A. Kl\"umper, A. Schadschneider, and J. Zittartz, 
Europhys. Lett. {\bf 24},  293  (1993);
B. Derrida, M.~R. Evans, V. Hakim, and V. Pasquier, J.~Phys.~A {\bf 26}, 
1493 (1993);
C. Lange, A. Kl\"umper, and J. Zittartz, Z.~Phys.~B {\bf 96},  267  (1994);
V. Karimipour and L. Memarzadeh, Phys. Rev.~B {\bf 77},  094416  (2008).

\bibitem{OestlundRommer95}
S. \"Ostlund and S. Rommer, Phys. Rev. Lett. {\bf 75}, 3537 (1995);
S. Rommer and S. \"Ostlund, Phys. Rev.~B {\bf 55}, 2164 (1997).

\bibitem{White92}
S.~R. White, Phys. Rev. Lett. {\bf 69}, 2863 (1992); 
Phys. Rev.~B {\bf 48}, 10345 (1993);
U. Schollw\"ock, Rev. Mod. Phys. {\bf 77}, 259 (2005);
R.~M. Noack and S.~R. Manmana, AIP Conf. Proc. 789, 93-163 (2005),
available at arXiv:cond-mat/0510321; 
K. Hallberg, Adv. Phys. {\bf 55}, 477 (2006).

\bibitem{Mcculloch07} See for example: I.~P. McCulloch, J. Stat. Mech. (2007)
  P10014, and references therein.

\bibitem{ReadSachdev91}
N. Read and S. Sachdev, Phys. Rev. Lett. {\bf 66},  1773  (1991);
S. Sachdev and N. Read, Int. J.~Mod. Phys.~B {\bf 5},  219  (1991), 
available at arXiv:cond-mat/0402109.

\bibitem{Sachdev92}
S. Sachdev, Phys. Rev.~B {\bf 45},  12377  (1992);
R. Tchernyshyov, R.~Moessner and S.~L. Sondhi, Europhys. Lett. {\bf 73},  278
  (2006);
M.~J. Lawler, L. Fritz, Y.~B. Kim, and S. Sachdev, Phys. Rev. Lett. {\bf 100},
  187201  (2008).

\bibitem{Vojta-00}
M. Vojta and S. Sachdev, Phys. Rev. Lett. {\bf 83}, 3916 (1999);
M. Vojta, Y. Zhang, and S. Sachdev, Phys. Rev.~B {\bf 62},  6721  (2000).

\bibitem{SachdevWang91}
S. Sachdev and Z. Wang, Phys. Rev.~B {\bf 43},  10229  (1991);
P. Nikoli\'{c} and S. Sachdev, Phys. Rev.~A {\bf 75},  033608  (2007);
M.~Y. Veillette, D.~E. Sheehy, and L. Radzihovsky, Phys. Rev.~A {\bf 75},
  043614  (2007).

\bibitem{VojtaRoesch08}
M. Vojta and O. R\"osch, Phys. Rev.~B {\bf 77}, 094504 (2008).

\bibitem{WeberVojta08}
H. Weber and M. Vojta, Phys. Rev.~B {\bf 77}, 125118 (2008).

\bibitem{Flint-07}
R. Flint, M. Dzero, and P. Coleman, arXiv:0710.1126v2[cond-mat.str-el];
arXiv:0710.1128v2 [cond-mat.str-el];
M. Dzero and P. Coleman, Physica~B {\bf 403},  955  (2008).

\bibitem{Wu-03}
C. Wu, J.-P. Hu, and S.-C. Zhang, Phys. Rev. Lett. {\bf 91}, 186402 (2003);
C. Wu, Mod. Phys. Lett.~B {\bf 20}, 1707 (2006).

\bibitem{Lecheminant-05}
P. Lecheminant, E. Boulat, and P. Azaria, Phys. Rev. Lett. {\bf 95}, 
240402 (2005);
C. Wu, Phys. Rev. Lett. {\bf 95}, 266404 (2005);
S. Capponi, G. Roux, P. Azaria, E. Boulat, and P. Lecheminant, Phys. Rev.~B 
{\bf 75}, 100503(R) (2007);
H. H. Tu, G.-M. Zhang, and L. Yu, Phys. Rev.~B {\bf 76}, 014438 (2007). 

\bibitem{Hamermesh89}
M. Hamermesh, {\em Group theory and its application to physical problems}
  (Dover, New York, 1989).

\bibitem{Cornwell84vol2}
J.~F. Cornwell, {\em Group theory in physics} (Academic Press, London, 1984),
  Vol.~II.

\bibitem{Arnold89}
V.~I. Arnold, {\em Mathematical Methods of Classical Mechanics} (Springer, New
  York, 1989).

\bibitem{MajumdarGhosh69jmp1}
C.~K. Majumdar and D.~K. Ghosh, J.~Math. Phys. {\bf 10},  1388  (1969);
J.~Math. Phys. {\bf 10},  1399  (1969);
P.~M. van~den Broek, Phys. Lett.~A {\bf 77},  261  (1980);
W.~J. Caspers, K.~M. Emmett, and W. Magnus, J.~Phys.~A: Math. Gen. {\bf 17},
  2687  (1984).

\bibitem{Holman69}
W.~J. Holman, J.~Math. Phys. {\bf 10},  1710  (1969).

\bibitem{Littelmann90}
P. Littelmann, J.~Alg. {\bf 130},  328  (1990).

\bibitem{Leung93}
E.~Y. Leung, J.~Phys.~A: Math. Gen. {\bf 26},  5851  (1993).

\bibitem{affleck:confinement} 
  I. Affleck, in {\em Dynamical Properties of Unconventional Magnetic
    Systems}, edited by A.~T. Skjeltorp and D. Sherrington (Kluwer
  Academic, Dordrecht, 1998), available at cond-mat/9705127; D.
  Augier, D. Poilblanc, E. S{\o}rensen, and I. Affleck, Phys. Rev.  B
  {\bf 58}, 9110 (1998); E. S{\o}rensen, I. Affleck, D. Augier, and D.
  Poilblanc, Phys. Rev. B {\bf 58}, R14701 (1998); M. Greiter, J.~Low
  Temp. Phys. {\bf 126}, 1029 (2002).

\bibitem{Greiter02prb1}
M. Greiter, Phys. Rev.~B {\bf 65},  134443  (2002);
Phys. Rev.~B {\bf 66},  054505  (2002).

\bibitem{Knabe88}
S. Knabe, J.~Stat. Phys. {\bf 52},  627  (1988).

\bibitem{NwachukuRashid76}
C.~O. Nwachuku and M.~A. Rashid, J.~Math. Phys. {\bf 17},  1611  (1976);
J.~Math. Phys. {\bf 18},  1387  (1977);
C.~O. Nwachuku, J.~Math. Phys. {\bf 20},  1260  (1979).

\end{thebibliography}
\end{document}